\documentclass[
    ,final            
  ]
  {aipproc}

\layoutstyle{6x9}

\begin{document}

\title{Time Domain Simulations of EMRIs using Finite Element Methods}

\classification{04.80.Nn,98.62.Js}
\keywords      {Extreme-mass-ratio inspirals, gravitational-wave source modeling}

\author{Carlos F. Sopuerta}{
  address={Institute for Gravitational Physics and Geometry,
           Center for Gravitational Wave Physics,    
           and Department of Astronomy \& Astrophysics,\\ 
           The Pennsylvania State University, University Park, PA 16802, USA}
}

\begin{abstract}
This is a brief report on time-domain numerical simulations of extreme-mass-ratio
binaries based on finite element methods.   We discuss a new technique for solving the
perturbative equations describing a point-like object orbiting a non-rotating massive
black hole and the prospects of using it for the evaluation of the gravitational self-force
responsible of the inspiral of these binary systems.  We also
discuss the perspectives of transferring this {\em technology} to the more astrophysically
relevant case of a central rotating massive black hole.
\end{abstract}

\maketitle


\section{Introduction} \label{intro}
Extreme-mass-ratio binaries in the inspiral stage of their evolution (EMRIs)
are considered to be a primary source of gravitational radiation to be detected 
by LISA.   They consist of a ``small'' object (SO), such a main sequence star, a stellar mass
black hole, or a neutron star orbiting a massive black hole (MBH), and the 
mass ratios of interest lie in the range $10^{-3}-10^{-6}\,$.  
Their study is expected to provide crucial information regarding MBHs at the center 
of galaxies, tests of the validity of general relativity, possibility of discerning 
among different theories of galaxy formation, etc.

Since the signal from EMRIs will be buried in the LISA {\em noise}, it is crucial to have a
good theoretical understanding of their evolution in order to produce accurate
waveform templates to be used in data analysis schemes for detection and extraction of physical
information.  There are several efforts in this direction, all based on perturbative methods
given the small mass ratios involved.  They differ mainly in the way in which
the {\em backreaction} on the SO is handled.  In order of
increasing complexity they are: (i) {\em Klugde} waveforms.  The concept is to use
approximations that allow quick generation of waveform templates~\cite{Babak:2006uv}.
(ii) The {\em adiabatic} approximation.  It is based on the idea that the long-term
evolution may be approximated by the {\em dissipative} part of the gravitational
{\em self-force}~\cite{adiabaticaprox}.
(iii)  The  self-force approach. It consists in evaluating the full self-force on the
SO~\cite{selfforce}, and from there to compute the modified motion and the
associated waveforms.
Approaches (i) and (ii) may provide useful templates for EMRI detection
but it is unlikely that they can be used to extract relevant physical information
(in particular the first multipole moments of the MBH).  
Moreover, it has been suggested~\cite{Pound:2005fs} that the adiabatic approximation may
not work as well as it was thought, in the sense that the conservative part 
of the self-force may have a significant effect in the long-term evolution.
Therefore, there is a good motivation to pursue the approach (iii).

The study of the dynamics of EMRIs via the self-force approach involves a number
of challenges~\cite{generalmaterial} and can be divided into the
following three stages: (A) The computation of the gravitational perturbations
produced  by the SO in the {\em background} spacetime of the MBH, in particular
at the SO location.  (B) Solving the equations of motion for the SO including its
own gravitational field [given by solving (A)].  (C) Extraction of meaningful
physical information, in particular the gravitational waveforms.   None of these parts has 
been completely solved.   The type of difficulties that appear depends
strongly on whether the SO is modeled as a point-like object or as an extended object.  
Here we adopt the first possibility, which is the one more commonly used.
Our discussion will focus on solving (A) with the perspective of using 
the result for (B).  This means to carry out (A) in a gauge suitable for the self-force
calculation.  For instance, in the Regge-Wheeler gauge, the reconstruction of metric perturbations
at the particle location involves singular terms.  On the contrary, the Lorentz 
gauge is a very convenient one since a scheme for regularizing the self-force,
the {\em mode-sum} scheme~\cite{Barack:2001gx}, has been formulated on it.

\section{Computing the perturbations produced by a point-like object using Finite Element Methods}
Point (A) consists in solving for the perturbations produced by a point-like object
orbiting a Kerr black hole (MBHs at the galactic centers are expected to have
high spins).  However, due to the complexity of the rotating case, it is advisable to deal first
with the case of a non-rotating Schwarzschild black hole.  We discuss later the case of Kerr.
In any case, in order to solve the corresponding perturbative equations  we need to
resort either to additional perturbative methods or to numerical techniques.    While
the first ones may have difficulties with orbits in the strong field regime, the second
ones should be capable of providing the solution for any type of orbit.  Among the
possible numerical methods, time-domain schemes seem better suited than frequency-domain ones.
In particular for highly eccentric orbits, the orbits with more astrophysical relevance,
for which the frequency-domain approach has more difficulties since one has to sum over
a large number of modes to obtain a good accuracy.

For a non-rotating MBH, the linear equations governing the perturbations can be decomposed 
in spherical harmonics.   Each harmonic decouples from the rest, and so do polar and axial modes.   
As a result each mode obeys inhomogeneous equations that involve only dependence on time and 
a {\em radial} coordinate $x$, i.e. they are one-dimensional partial differential equations (PDEs).
Of crucial importance is the structure of the sources of these inhomogeneous equations.
They are generated by the energy-momentum distribution of the particle and
can be represented as follows
\begin{equation}
{\cal S}^{\ell m} = F^{\ell m}_0\, \delta\left[x-x^{}_p(t)\right] +
F^{\ell m}_1\, \delta'\left[x-x^{}_p(t)\right]+
F^{\ell m}_2\, \delta''\left[x-x^{}_p(t)\right]\,, \label{sourceterms}
\end{equation}
where $x^{}_p(t)$ is the radial motion of the particle,
$(F^{\ell m}_0\,,\,F^{\ell m}_1\,,\,F^{\ell m}_2)$ are functions of $t$ and $x$,
and $\delta$, $\delta'$, and $\delta''$ denote the Dirac delta distribution and
its first and second derivatives respectively.  When the perturbations are described
by pure metric perturbations, like in the Lorentz gauge, we have $F^{\ell m}_1 =
F^{\ell m}_2 = 0$.  When they are described by variables of the type of the 
Regge-Wheeler and Zerilli master functions only $F^{\ell m}_2 = 0$.  Finally,
if the perturbations are described by curvature-type variables we have all the
terms.  This representation of the source terms shows their singular structure.
In the best case, the solution of the perturbative equations will be continuous
but with discontinuous first radial derivatives.  In the worst case ($F^{\ell m}_2 \neq 0$),
the solution will diverge as we approach the particle location.

When solving numerically the perturbative equations it is then crucial that we 
discretize appropriately the source terms.  This is a non-trivial task but there are 
some prescriptions that produce reliable results.  As a matter of fact,  these prescriptions 
are based on integral forms of the equations, which allow the use of the known properties of
the Dirac delta distribution, avoiding the introduction of {\em artificial} 
regularizations of it like a Gaussian packet.  
The first such prescription was proposed in~\cite{Lousto:1997wf} and is based on a finite 
differences scheme that resembles the procedures used to derive finite volume algorithms.
Its main drawback is that it cannot be easily generalized to higher-dimensional domains, 
and hence to the Kerr case.

We have recently proposed~\cite{Sopuerta:2005gz} a different prescription 
based on FE methods for the spatial discretization.  In brief 
(see details in~\cite{Sopuerta:2005gz,Sopuerta:2005rd}),
since the spatial domain is one-dimensional (the radial direction), the domain discretization consists
in a division into disjoint intervals, our {\em elements}.
Then, we assign to each element (interval) a finite-dimensional functional space 
in order to approximate locally the solution of our equations.  Typically, these functional spaces 
are made out of polynomials, and the accuracy and convergence properties
of the resulting FE algorithm depend strongly on how we choose them.  
We can approximate the solution of our PDEs, say $\psi$, by an expansion in nodal functions,
$n^{}_i$ [$n^{}_i(x^{}_j) = \delta^{}_{ij}$], constructed from the functional spaces:
\begin{equation}
\psi^{}_h(t,x) = \sum_{i=0}^{N}\psi^{}_i(t)  n^{}_i(x) \,. \label{feaprox}
\end{equation}
Then, one transforms the PDEs into their
{\em weak} form, an integral form that consists in multiplying the equations by
an arbitrary test function, integrating over the spatial domain, and applying integration
by parts to eliminate second spatial derivatives while incorporating non-essential
boundary conditions (for instance, Sommerfeld boundary conditions).  We denote the
weak form of the equations by ${\cal E}[\phi,\psi] = 0$, where $\phi$ is the test 
function.  In a Galerkin-type FE formulation, the discretized
equations are obtained by imposing the vanishing of all the {\em residuals}:
\begin{equation}
{\cal E}^{}_i \equiv {\cal E}[n^{}_i, \psi^{}_h] = 0~~~(i=0,\ldots,N)\,. \label{residuals}
\end{equation}
Introducing~(\ref{feaprox}) into~(\ref{residuals}) we get as many equations as independent
functions $\psi^{}_i(t)$ we have in~(\ref{feaprox}).  These equations are ordinary
differential equations that can be solved by using adequate integrators.
The important point for our discussion is that the equations that we obtain from~(\ref{residuals})
will contain terms of the type 
\begin{equation}
\int dx\; n^{}_i(x)\,{\cal S}^{\ell m}\,,
\end{equation}
which can be easily evaluated by using the properties of the Dirac delta distribution.
And this is essentially how the FE method provides a prescription for the
singular source terms in~(\ref{sourceterms}).  This prescription depends on
our choice of nodal functions, and hence on the domain discretization.
There are two important factors: (i) The degree of the polynomials of the functional spaces
of the FE discretization.  (ii) The location of the particle in the mesh.  Essentially 
whether it is located at a node or inside an element.

The degree of the polynomials determines the degree of convergence of the FE method.
For instance, for linear elements ($n^{}_i\sim ax+b$) and smooth functions we expected
second-order convergence.  However, the singular structure of our source terms may
deteriorate it.  Actually, linear elements will not be able to resolve the $\delta''$ 
distribution since the action of it on any linear element will give zero.  As a consequence,
if we want to evaluate accurately self-forces from the perturbations, we need to choose 
carefully the type of elements.

The location of the particle plays a role in the following sense:  If the particle
is located inside an element, then the integrals of the source terms are evaluated
using the properties of the Dirac delta distributions.  This is simple in the sense
that we only need to know the particular element that contains the particle.  However,
the convergence rate may be lower than expected, which happens in the case where
$F^{\ell m}_1\neq 0\,.$  On the contrary, if we locate the particle in a node, we
can assign two values of the perturbative variables (and their derivatives depending
on the exact singular structure of the source terms) to this node and impose there
the {\em jumps} in the derivatives of these variables or the variables themselves.
These jumps can be analytically derived from the perturbative equations~\cite{Sopuerta:2005gz}.
The advantage of this method is that it preserves the expected convergence rate
of the FE method.  The drawbacks are that implicit evolution algorithms may be subject
to a Courant-Friedrichs-Lax stability condition, and that in order to keep the particle
always at a node we need to implement moving mesh techniques.

We have performed simulations using the techniques described here, using also refinement
of the meshes, in the Regge-Wheeler gauge~\cite{Sopuerta:2005gz} (in this case
$F^{\ell m}_2 = 0$).  We found an excellent
agreement with other previous calculations of the energy and angular momentum luminosities
for all kinds of orbits, including highly eccentric ones.  Regarding the different numerical
implementations discussed above, we found that their performance depends strongly on
the type of orbit that we are considering, in the sense that the {\em best} scheme changes 
among the different types of orbits.  We also found that the mesh refinement and moving
mesh techniques help substantially in reducing the computational resources used by the
simulations without damaging the accuracy.   On the other hand, we have also performed simulations in the Lorentz gauge
(in this case $F^{\ell m}_1 = F^{\ell m}_2 = 0$), which is a suitable one for self-force
computations.  So far, the only known results in this gauge are only for circular orbits and
can be found in~\cite{Barack:2005nr}.  We have been able to perform simulations for all kinds of 
orbits and the results will be published elsewhere.

\section{Extension to the case of a rotating MBH}
One of the advantages of the FE method is that many of the techniques discussed above for
the case of Schwarzschild can be transfered to the case of Kerr, in contrast with previous
approaches.  The PDEs governing the perturbations of Kerr can be two- or three-dimensional, 
depending on whether we factor out the dependence on the azimuthal angle.  In any case, 
prescriptions for the regularization of the corresponding source terms can be derived in exactly 
the same way as in the Schwarzschild case if the particle is always inside an element.  
In contrast, in order to use the techniques corresponding to the case
in which the particle is at a node we would need to derive a new framework.
An important point to be taken into account is the fact that the solutions of the perturbative 
equations, in the case in which we separate the azimuthal dependence, diverge logarithmically as we approach
the particle.  This fact will require either the substraction {\em a priori} of the
singular part of the solution or a careful numerical regularization.

Another important feature of the FE method that may be of crucial importance for the
rotating case is the natural way in which refinement schemes can be implemented.
Taking into account that we will be dealing either with two- or three-dimensional
domains, refinement may be a necessity in order to carry out simulations at the 
required accuracy.   Here, it is important to mention that, to a certain extent,
we can import  to the Kerr case the refinement schemes used in the Schwarzschild case.
The way of doing this is to use for the Kerr case quadrilateral/hexahedral elements,
since for these elements some classes of nodal functions can just be obtained from the 
one-dimensional ones by means of {\em tensorial} products.

In summary, the main challenges for the Kerr case seem to be the
substraction of the logarithmic singularities and the implementation of an efficient
mesh refinement scheme.  The rest of ingredients in the simulations do not seem
to present additional difficulties with respect to the non-rotating
case, in which we have already acquired a considerable amount of experience.


\begin{theacknowledgments}
We acknowledge the support of the Center for Gravitational Wave Physics funded by the 
National Science Foundation under Cooperative Agreement PHY-0114375, and partial
support from NSF grant PHY0244788.
\end{theacknowledgments}

\bibliographystyle{aipproc}   

\bibliography{references}

\begin{thebibliography}{14}
\expandafter\ifx\csname natexlab\endcsname\relax\def\natexlab#1{#1}\fi
\providecommand{\enquote}[1]{``#1''}
\expandafter\ifx\csname url\endcsname\relax
  \def\url#1{\texttt{#1}}\fi
\expandafter\ifx\csname urlprefix\endcsname\relax\def\urlprefix{URL }\fi
\providecommand{\eprint}[2][]{\url{#2}}

\bibitem{Babak:2006uv}
S.~Babak, H.~Fang, J.~R. Gair, K.~Glampedakis, and S.~A. Hughes  (2006),
  \eprint{gr-qc/0607007}.

\bibitem{adiabaticaprox}
Y.~Mino, \emph{Phys. Rev.} \textbf{D67}, 084027 (2003);
S.~A. Hughes, S.~Drasco, E.~E. Flanagan, and J.~Franklin, \emph{Phys. Rev.
  Lett.} \textbf{94}, 221101 (2005);
N.~Sago, T.~Tanaka, W.~Hikida, K.~Ganz, and H.~Nakano, \emph{Prog. Theor.
  Phys.} \textbf{115}, 873--907 (2006).

\bibitem{selfforce}
Y.~Mino, M.~Sasaki, and T.~Tanaka, \emph{Phys. Rev.} \textbf{D55}, 3457--3476
  (1997);
T.~C. Quinn, and R.~M. Wald, \emph{Phys. Rev.} \textbf{D56}, 3381--3394 (1997).

\bibitem{Pound:2005fs}
A.~Pound, E.~Poisson, and B.~G. Nickel, \emph{Phys. Rev.} \textbf{D72}, 124001
  (2005).

\bibitem{generalmaterial}
E.~Poisson, \emph{Living Rev. Relativity} \textbf{7}, 6 (2004);
C.~O. Lousto, editor, \emph{Gravitational Radiation from Binary Black Holes:
  Advances in the Perturbative Approach}, vol.~22, \emph{Class. Quant.
  Grav.}, Institute of Physics, London, 2005.

\bibitem{Barack:2001gx}
L.~Barack, Y.~Mino, H.~Nakano, A.~Ori, and M.~Sasaki, \emph{Phys. Rev. Lett.}
  \textbf{88}, 091101 (2002).

\bibitem{Lousto:1997wf}
C.~O. Lousto, and R.~H. Price, \emph{Phys. Rev.} \textbf{D56}, 6439--6457
  (1997).

\bibitem{Sopuerta:2005gz}
C.~F. Sopuerta, and P.~Laguna, \emph{Phys. Rev.} \textbf{D73}, 044028 (2006).

\bibitem{Sopuerta:2005rd}
C.~F. Sopuerta, P.~Sun, P.~Laguna, and J.~Xu, \emph{Class. Quant.
  Grav.} \textbf{23}, 251--285 (2006).

\bibitem{Barack:2005nr}
L.~Barack, and C.~O. Lousto, \emph{Phys. Rev.} \textbf{D72}, 104026 (2005).

\end{thebibliography}

\end{document}